# *Optical control of one and two hole spins in interacting quantum dots*

Alex Greilich, Samuel G. Carter, Danny Kim, Allan S. Bracker and Daniel Gammon

Naval Research Laboratory, Washington, DC 20375, USA

**A single hole spin in a semiconductor quantum dot has emerged as a quantum bit that is potentially superior to an electron spin. A key feature of holes is that they have a greatly reduced hyperfine interaction with nuclear spins, which is one of the biggest difficulties in working with an electron spin. It is now essential to show that holes are viable for quantum information processing by demonstrating fast quantum gates and scalability. To this end we have developed InAs/GaAs quantum dots coupled through coherent tunneling and charged with controlled numbers of holes. We report fast, single qubit gates using a sequence of short laser pulses. We then take the important next step toward scalability of quantum information by optically controlling two interacting hole spins in separate dots.**

A semiconductor qubit offers powerful advantages for quantum information, including a scalable solid state platform, integration with electronic or photonic circuits [1], ultrafast optical control [2,3,4,5,6,7,8], and relatively long lifetimes [9,10,11]. Until recently, most efforts in semiconductors have utilized the spin of the electron as a qubit. However, a major difficulty of using electron spins is the strong hyperfine interaction, which couples the qubit to a bath of nuclear spins [12]. Random fluctuations of the nuclear spin polarization lead to electron spin dephasing. In the self-assembled InAs/GaAs quantum dots (QDs) discussed in this article, the dephasing time $T_2^*$ is typically 1-2 ns [6,13], while in larger, electrically-defined or interface fluctuation QDs, $T_2^*$ is 5-10 ns [2,14]. This nuclear spin problem has motivated studies of hole spins, which have an order of magnitude weaker hyperfine interaction [15,16,17,18,19]. Early optical studies have been quite promising, giving long spin relaxation ($T_1 \sim 1$ ms) [9, 10] and dephasing times ($T_2^* > 100$ ns) [11] in InAs self-assembled QDs.

While a QD hole can be treated as a pseudo-spin qubit and optically measured and controlled like an electron, in many ways its properties are unique and interesting. A hole is the absence of an electron in the otherwise full valence band states. In the p-type valence states, spin-orbit interaction is strong, which profoundly changes the nature of the hole spin and makes its properties sensitive to electric fields, strain, and variations in the confinement potential [20]. In particular, anisotropy in the confinement potential in self-assembled QDs (stronger confinement along the growth direction than the transverse direction) gives significant anisotropy in the spin properties, such as the Zeeman interaction [21]. Similarly, the hyperfine interaction for a hole is highly anisotropic, with an approximately Ising-like ($AS_z I_z$) form [15]. This is predicted to greatly reduce dephasing in a transverse magnetic field [16], even beyond the order-of-magnitude naively expected, which could explain the long dephasing time that has been reported [11].

Moreover, hole tunneling between two such QDs exhibits new phenomena such as electric field resonances of the g-factor [22], which can be used to control the spin for quantum operations [23], and reversal of the bonding and anti-bonding states [24]. Remarkably, and in contrast to the other spin interactions, the exchange interaction between two hole spins in two QDs is predicted to be approximately isotropic or Heisenberg-like ($J\vec{S}_1 \cdot \vec{S}_2$) [25]. As a result, the two-spin states should separate into a singlet and a set of degenerate (or nearly degenerate) triplets, and ideas for optically controlling entanglement of two spins developed for two electrons can be brought over to two holes [8].

With these differences between the spin of an electron and hole in mind, quantum control demonstrations are now needed to establish the hole's viability as a qubit. We take this step by demonstrating ultrafast optical control of hole spins in a system of coupled QDs for three different regimes: (i) an effectively isolated single hole spin, (ii) weakly interacting hole spins, and (iii) strongly interacting hole spins. These results are quite promising for hole spin qubits and reveal future opportunities and challenges.

**Results**

**Holes in two tunnel-coupled QDs.** We have developed a structure in which two vertically stacked quantum dots were grown within a p-type Schottky diode structure whereby an applied voltage permits holes to be injected one by one into each of the QDs (Fig. 1). To characterize the system, optical absorption and emission are measured through *cw* resonant transmission and photoluminescence spectroscopy, respectively (Fig. 2). Both the configuration of holes and their corresponding optical transitions to excited states ($X^+$, $X^{2+}$, $X^{3+}$) are determined from the optical spectra. With increasing bias (left to right in Fig. 2a and 2b), the stable hole configurations in the two QDs go from (0,1) to (1,1) to (1,2) to (2,1), with vertical dashed lines in the figures separating the configurations. The numbers in parentheses correspond to the number of ground state holes in each QD ("red", "blue"). The optical transition for each of these charge configurations involves the excitation or recombination of an additional electron-hole pair in the red QD as shown at the top of Fig. 2b. The blue transition is ~30 meV higher in energy. The optical resonance frequency shifts for each of the charge configurations because of differing Coulomb energies [26].

Anti-crossings of the spectral lines are a signature of coherent tunneling between the two QDs [27,28,29]. In particular, an anti-crossing structure at -0.52 V to the left of the stable (1,1) voltage region corresponds to the exchange splitting between the singlet-triplet spin states [8]. This exchange splitting gradually decreases with increasing bias and in this sample is not resolved in the voltage region where the (1,1) configuration is stable. A model calculation (fit to the PL data) gives an exchange interaction of ~1 GHz, which is less than the optical linewidths (~2 GHz). We have also studied a second sample with a thinner tunnel barrier with a much larger exchange energy of ~20 GHz. The two-hole configuration (1,1) is of special interest because it permits us to demonstrate coherent optical control of two interacting hole spins.

**Optical control of a single hole spin.** We first demonstrate ultrafast optical control of a single hole spin, which has not previously been achieved. We can do this in the same quantum dot structure simply by changing the voltage bias to the (1,2) configuration. By maintaining two spin-paired holes in the blue dot, the only degree of freedom is the single hole spin in the red dot (see Fig. 1b). In a transverse magnetic field ($B_x = 6$ T), the two spin configurations of the (1,2) state and an optically excited state, $X^{3+}$, provide a three-level Λ system by which single spin initialization and rotation can be performed.

For our quantum control demonstration we apply a Ramsey interference technique to show control of the hole spin (Fig. 1a) [3, 5, 7, 8]. A 20 ns long pulse from a narrowband diode laser was used to prepare the spin of the hole by pumping one transition of the Λ-system (Fig. 3a) [10, 30, 31, 32, 33]. During this pulse the hole spin is initialized to the ⇑ eigenstate of the system parallel to the magnetic field. This initial state is represented as the south pole of the Bloch sphere in Fig. 3b. Next, the Ramsey pulse sequence is applied. For this, we use two circularly polarized pulses, each of which rotates the spin about the optical axis $k_z$ by $\pi/2$ (see the red arrows in Fig. 3b) [2, 3, 4, 5]. Both pulses were 13 ps long (bandwidth = 150 μeV) and detuned 300 μeV below the low-energy transition to avoid real excitations. After the first pulse the hole spin is driven to the equator of the sphere where it starts to precess around the vertical axis, which is defined by the direction of the applied magnetic field, $B_x$. After a delay $\tau$ the second pulse rotates the spin vector toward the north or south pole, depending on the phase, thus converting the phase into a population. This population is determined by measuring small changes in the transmission of the diode laser pulse. This initialization/rotation/measurement cycle is repeated with a period ten times that of the repetition period of the laser ($T_{Cycle} = 10\ T_R = 123$ ns) and an averaged signal is plotted in Fig. 3c.

From the Ramsey fringe curve we can obtain the dephasing time of the single-hole spin. The hole spin coherence persists with little decay for the first 5 ns (The weak signal near zero delay will be discussed later). We extended the delay by 1 and then 2 laser repetition periods of 12.3 ns to measure the decay of the phase oscillations. Figure 3d shows the corresponding signal fitted to a Gaussian envelope that decays with an inhomogeneous dephasing time $T_2^* = 20.7$ ns. This value is an order of magnitude larger than that measured for electrons in similar studies of self-assembled In(Ga)As QDs [6, 13]. In these earlier studies dephasing was limited by the interaction of the electron spin with the fluctuating nuclear environment. While the hyperfine interaction may induce some dephasing here, we find that $T_2^*$ is instead limited by electrical-noise-induced fluctuations in the hole g-factor. That is, because the hole precession frequency varies with voltage, having a slope of ~10 MHz/mV, fluctuations in the electrical potential lead to fluctuations in the precession frequency. In this sample we have voltage fluctuations of a few millivolts (see Methods), which should give a $T_2^*$ of tens of nanoseconds – consistent with the measured value. This result demonstrates that the hyperfine contribution to $T_2^*$ is at least an order of magnitude weaker than for electrons, consistent with other studies that found a reduced interaction strength [17, 18, 19]. A previous experiment has reported a $T_2^*$ greater than 100 ns, using frequency domain measurements. This longer time may indicate that electrical-noise-induced fluctuations of the g-factor were smaller in that sample, perhaps because of less impurity-induced electrical noise.

Although hyperfine interactions are not dominant in the dephasing of the hole spin, there is evidence that they are involved in its evolution. This is seen in the saw-tooth form of the fringe pattern in Fig. 3c (see the inset of Fig. 3e). This behavior was previously observed in the Ramsey fringes of an electron spin, and was explained by nuclear spin dragging [34]. The origin is likely the same here (see Methods). This effect is seen also through the Fourier transform of the signal (Fig. 3e). The main peak at 7 GHz corresponds to Larmor precession of the hole spin in the transverse magnetic field of $B_x = 6$ T, giving a hole g-factor in the red QD of 0.084. The higher harmonics are the result of the saw-tooth form of the time-domain signal. Line following and hysteresis in the *cw* transmission spectra are also observed here – an effect that has been previously shown to arise from feedback with the nuclear system in *cw* studies of electron systems [35, 36]. We suspect that the small signal for short delays is due to these nuclear dragging effects, which can shift the transition out-of-resonance with the laser.

Full control of the hole spin, *i.e.* rotation about an arbitrary axis by any angle, is achieved by a combination of optical pulse power, which rotates the spin up the Bloch sphere, and time delay, which allows precession around the sphere [3]. To demonstrate we measure the amplitude of the Ramsey fringes as a function of pulse power as displayed in Fig. 3f. The first maximum occurs for two π/2 pulses and the first minimum for two π pulses. These oscillations (essentially spin Rabi oscillations) confirm that this pulse sequence can optically rotate the hole spin to any point on the Bloch sphere and thus satisfy the requirements for qubit implementation. We find that the spin rotation angle has a sublinear dependence on power ($\sim p^{0.78}$). This behavior is due to insufficient pulse detuning for adiabatic elimination of the exciton state as reported in previous studies of electron spin [3, 8].

**Optical control of two interacting hole spins: weak interaction.** For this experiment the voltage bias on the sample is adjusted to give the stable two-hole configuration, (1,1), where the holes are in separate dots. Again we directly control and measure the spin in the red QD with near-resonant optical pulses. Since the exchange splitting is not larger than the optical transition linewidths in the stable region of this sample (Fig. 2), transitions in the red QD are not sensitive to the spin state in the blue QD. In this limit, we can draw the level diagram in Fig. 4d as two Λ-systems, differing only in the spin state of the blue QD. The initialization/measurement laser pumps two transitions (solid arrows) that are not resolved, initializing into an incoherent mixture of ⇑⇓ and ⇑⇑. The short pulses also act on both Λ-systems, performing spin rotations on the red QD spin, regardless of the blue QD spin. The hole spin in the blue QD is not directly controlled but it interacts with the spin in the red QD through the exchange interaction $J$. The resulting Ramsey fringe signal is shown in Fig. 4a. There is a clear beating in the data that depends on the voltage. In the Fourier transform it is seen that the peak at the Larmor spin resonance has split into two with a splitting given by the exchange frequency (Fig. 4b). In addition, there are additional peaks at sums and differences of these primary resonance frequencies that come from nonlinearities arising from interactions with the nuclear spin, as in the single spin case.

We identify the beating in the Ramsey fringes as tunneling-induced exchange between the QD spins, which is tuned with applied voltage. If the blue dot has spin up (down) the precession frequency of the red dot spin is shifted up (down) by $J/2$. Since the hole spin in the blue QD is not initialized up or down, averaging over these two possible initial conditions gives a beating at $J$. The voltage dependence of the beating frequency (Fig. 4c) is consistent with the exchange frequency deduced from the photoluminescence spectrum. In addition, with decreasing magnetic field we find that the Larmor frequency (average of $f_1$ and $f_2$) decreases but the beating frequency (difference between $f_1$ and $f_2$) does not, as expected. Finally we note that the measured $T_2^*$ for the coupled hole spins (21 ns) is approximately the same as that for the single hole spin, again because of electric field fluctuations of the g-factor.

**Optical control of two interacting hole spins: strong interaction.** With the first sample we are in a regime of small exchange coupling. With the second sample we increase the exchange interaction by an order of magnitude to provide distinct singlet and triplet lines in the stable region of the charging sequence. A laser field resonant with the red QD can then be used to initialize to the singlet state. Fig. 5a displays a photoluminescence map of this sample at zero magnetic field. The two lower energy lines correspond to emission from the $X^{2+}$ exciton states to the singlet and triplet spin states of the (1,1) charge configuration.

The level diagram for the (1,1) configuration is displayed in Fig. 5b, where the eigenstates are now written in the singlet-triplet basis. The singlet, $S=(|\Uparrow\Downarrow\rangle - |\Downarrow\Uparrow\rangle)/\sqrt{2}$, and $m_s=0$ triplet, $T_0=(|\Uparrow\Downarrow\rangle + |\Downarrow\Uparrow\rangle)/\sqrt{2}$, are part of two lambda systems that differ in the helicity of their circular polarization selection rules. The other triplets, $T_+=|\Uparrow\Uparrow\rangle$ and $T_-=|\Downarrow\Downarrow\rangle$ are nominally not connected to $S$ and $T_0$ through optical transitions. However, we find that optical pumping can deplete the $T_\pm$ states, most likely due to hyperfine coupling of electron or hole spins that mixes the $X^{2+}$ states or triplet states, respectively. With the laser resonant with the triplets, we can thus initialize the system into the singlet state, in which the hole spins are entangled [8].

Ramsey fringe experiments were performed in this system, as shown in Fig. 5c at a bias of -0.6V and zero magnetic field. The optical pulses are short enough (~13 ps) compared to the exchange interaction that they still act as single qubit rotations on the red QD. On the $S$-$T_0$ Bloch sphere (see inset), the pulses rotate $S$ toward $T_0$ [8]. Initializing to the $S$ state with a cw laser on the triplets, the first pulse rotates the Bloch vector to the equator, generating a superposition of $S$ and $T_0$ that evolves in time. The Ramsey fringes observed here, in contrast to those presented previously in this article, are not due to precession in an external magnetic field. Instead, they are entirely due to the interaction between hole spins.

The frequency of the Ramsey fringes is about 20 GHz, consistent with the $S$-$T$ splitting in Fig. 5a at this bias. However, there is a beating in the Ramsey fringes at a frequency of 1.9 GHz. We assert that this beating is due to a zero-field splitting between $T_\pm$ and $T_0$. One frequency is due to a $S$-$T_0$ superposition and one due to a $S$-$T_\pm$ superposition. This splitting likely arises from an anisotropic contribution of ~10% to the isotropic Heisenberg exchange. This anisotropic exchange is another manifestation of the stronger spin-orbit character of the holes. With this assumption we are able to model well the measured fringe pattern as shown in Fig. 5c.

The fringe pattern decays significantly faster in this second sample ($T_2^* < 600$ ps), again because of electrical noise. However, now it is because the exchange splitting is a strong function of voltage. The frequencies of the Ramsey fringes are plotted in Fig. 1d as a function of voltage, with the measured $S$-$T$ splitting plotted for comparison. The large variation of the exchange frequency with bias (70 MHz/mV at -0.6V) means the system is sensitive to electrical noise. Indeed, we find that $T_2^*$ gets shorter at higher voltages as this slope increases, indicating that it plays a major role in dephasing. This dephasing could be minimized by optimizing the sample structure to shift the stability range of the (1,1) configuration to the minimum of the *S-T* splitting, where the slope is zero.

**Discussion**

We have presented fast, coherent rotations of a single hole spin, and also the first demonstration of two interacting hole spins in two QDs. The dephasing time of the hole was an order of magnitude longer than that reported for a single electron in a similar QD system with similar measurements [6, 13]. Instead of being limited by the hyperfine interaction, the hole spin dephasing time is primarily determined by electrical-noise fluctuations that affect the spin through the spin-orbit interaction. We expect dramatic improvements in the dephasing time by reducing electrical-noise fluctuations or by engineering the system to be less sensitive. Also, because these environmental interactions typically evolve much more slowly than the hole spin system, spin echo techniques should lead to much longer lifetimes as shown recently for electron spins [6]. In fact, preliminary spin echo measurements show significant enhancements in the dephasing time of the hole spin.

Controlling the interaction between spins is also essential as it is the basis for two-qubit gates. We have succeeded in generating a tunable interaction between spins of tens of GHz, orders of magnitude higher than in any other system, while retaining single qubit control using short optical pulses. The interaction is also largely isotropic, and sensitivity to electric fields can clearly be substantially improved. In the current systems the interaction between spins is tunable but always on, which may make it challenging to realize controllable interactions between many qubits. One advantage of hole spin qubits is that we can dramatically reduce the tunneling interaction of holes in the ground state, due to the large effective mass of holes [37]. The interaction can instead be produced in the excited states through electron tunneling or exciton-exciton interactions, giving ultrafast optical pulse control of the interaction [38, 39, 40].

**Methods**

**Sample structure.** Two vertically stacked InAs dots were grown by molecular beam epitaxy on a p+ GaAs substrate and embedded in a GaAs Schottky diode. The two samples consists of: a p-type Be-doped buffer layer; a 25nm GaAs spacer barrier; two InAs dot layers of 2.8 and 3.2nm separated by a 4nm or 6nm GaAs tunnel barrier; and a 280nm GaAs capping layer. The top dot was grown thicker than the bottom dot such that it has the lower energy optical transition, and such that only holes coherently tunnel between the QDs. The absorption linewidths in the samples are measured to increase with the slope of the Stark shift, which we attribute to electrical fluctuations that give rise to fluctuations in the transition energy. A voltage fluctuation of 4.3 mV was derived from the linear fit to the linewidth vs. slope of the Stark shift for the $X^{3+}$ transition.

**Measurement techniques.** The sample surfaces were covered with a 5 nm layer of Ti and a 120 nm thick Al layer, which was patterned with 1µm apertures. The sample was placed on piezo-stages in a He-flow cryostat at 5 K. 0.68 NA aspheric lens was placed in the cryostat to focus the laser on the aperture and also to collect the light for the photoluminescence measurements. Light transmitted through the aperture was focused onto an avalanche photodiode. The resonant transmission signal was probed by means of Stark-shift modulation spectroscopy using narrow-linewidth *cw* laser (nanoelectronvolt range) and lock-in techniques. The sample was modulated with a square-wave voltage at 10 kHz superimposed on the fixed d.c. voltage for phase sensitive detection of the transmitted intensity. For the Ramsey fringe experiments, a Titanium:sapphire laser was used with 13 ps pulses and repetition period of 12.3 ns. Circular polarization of pulses was opposite to that of the *cw* probe and was rejected using a polarizer. Pulses were detuned below the low energy exciton transition by 200-300 µeV.

**Nuclear hysteresis effects.** The saw tooth behavior observed in the Ramsey fringes can be understood essentially the same as that explained for electron spins in Ref. 34. The basic idea is that the rate of nuclear spin flips is proportional to the excitation probability by the initialization/measurement pulse, which depends on the precession phase at the second pulse. As in Ref. 41, a random walk of the nuclear polarization changes the precession frequency $\omega$ until the commensurate condition, $\omega\tau = 2\pi n$, is satisfied, at which no excitation or nuclear spin flips occur. As the pulse delay is slowly varied, the nuclear polarization adjusts to continue satisfying this condition but jumps when it pushes the exciton transition out of resonance with the laser. This gives rise to the nonlinear saw-tooth behavior observed. We also suspect that the small signal for small delays is due to the fact that the nuclear polarization required to satisfy the commensurate condition is much larger for small $\tau$, which tends to shift the transition out-of-resonance with the laser.

**Interacting spin states.** For the (1,1) hole configuration the ground state spin eigenstates at zero magnetic field are the singlet, $S = (\Uparrow\Downarrow - \Downarrow\Uparrow)/\sqrt{2}$, and three triplets, $T_0 = (\Uparrow\Downarrow + \Downarrow\Uparrow)/\sqrt{2}$, $T_- = \Downarrow\Downarrow$, and $T_+ = \Uparrow\Uparrow$. At high magnetic fields we expect a mixing of $S$ and $T_0$ due to a difference in g-factors $\Delta g$ between the red and blue QDs. The g-factor is sensitive to valence band mixing, which can vary from dot to dot and with orbital wavefunction changes. A significant difference of $\Delta g/g = 42\%$ has been measured in the Faraday field geometry for the 6 nm barrier sample, and we expect a similar relative difference in the Voigt geometry. For $\Delta g \mu B \gg J$, $S$ and $T_0$ are mixed to give eigenstates of $\Uparrow\Downarrow + \delta \Downarrow\Uparrow$ and $\Downarrow\Uparrow - \delta \Uparrow\Downarrow$, where $2\delta = J/\Delta g\mu B \sim 0.4$. This limit is reasonable for the weakly-interacting 6 nm barrier sample, and we can represent the ground states and exciton states as shown in Fig. 4d, where $\delta$ terms have been omitted for simplicity. We also note that the red dot precession frequency changes between the (1,2) and (1,1) configurations, going from 7 GHz to 5.5 GHz (average of the two frequencies, $f_1$ and $f_2$). This change is presumably due to a change in the orbital wavefunction in the red dot as number of holes in blue dot changes from two to one.

In the strongly-interacting 4 nm barrier sample, the singlet-triplet states should be good eigenstates since the exchange interaction is large and experiments are performed at zero magnetic field. The beating in the Ramsey fringes can be explained by anisotropy in the exchange interaction. For a Hamiltonian $H = J\vec{S}^R \cdot \vec{S}^B + 2J'S_z^R S_z^B$, the isotropic $J$ term gives rise to the singlet-triplet splitting, and the $J'$ term splits $T_0$ away from $T_+$ and $T_-$ by $J'$. ($S_z^R$ and $S_z^B$ are spin operators for the red and blue QDs, respectively, with eigenvalues $\pm 1/2$.) The anisotropy axis, $z$ in this Hamiltonian, determines the basis of the eigenstates. Our model of the Ramsey fringes gives a beating only when the anisotropy axis is in-between the growth direction and the plane of the QD. In this basis, the optical pulse rotates the $S$ state to a superposition of all four basis states, giving rise to the different frequencies observed. The selection rules in Fig. 5b will also be modified in this basis, allowing diagonal transitions. The origin of the anisotropy and its orientation likely arises from the spin-orbit interaction combined with structural asymmetry, such as a relative lateral displacement of the two QDs [42].

**Fidelity estimates.** At B = 0 T no spin pumping was observed in the 6 nm barrier quantum dot sample. With higher magnetic fields the pumping was recovered, reaching the initialization fidelity of 91% at $B_x = 6$ T as estimated from the bleaching of the higher energy transition in the single-laser differential transmission. To calculate this we compare the bleached signal level with the signal where pumping is suppressed by a single detuned π-pulse. The fidelity is then $F_{Pump} = 1-\alpha/\alpha_0 = 0.91$, where $\alpha$ is the signal level at 6 T and $\alpha_0$ is the signal level with the π-pulse. We estimate the fidelity of single-qubit gates using data from Fig. 3f, which plots the Ramsey fringe amplitude as a function of pulse power. The oscillations show decay that we attribute to pulse-induced decoherence. From the fit to a decaying sine, we determine the expected amplitude for two π/2-pulses with no dephasing, $A_{fit}$, and compare this with the maximum signal at 0.66 µW, $A_{meas}$. If each pulse reduces the Bloch vector by a factor D, then $D^2 = A_{meas}/A_{fit} = 0.96$, giving a gate fidelity of $F = (1+D)/2 = 0.99$ (Ref. 3).

**Acknowledgements** This work was supported by NSA, ARO MURI, and ONR.

Correspondence and requests for materials should be addressed to DG (gammon@nrl.navy.mil).

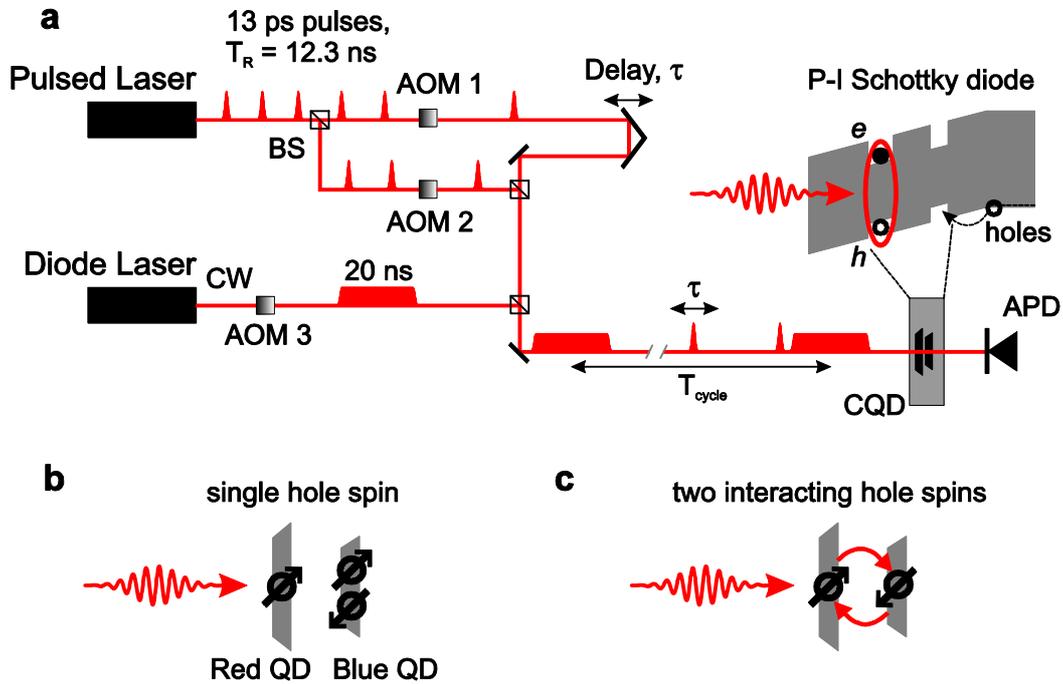

**Figure 1 | Design of experiment. a**, Setup configuration for Ramsey interference measurement with continuous wave diode laser used to initialize and read out the spin polarization, and pulsed laser to induce the spin rotation. $T_R$ – repetition period of the laser, BS – non-polarizing beam splitter, AOM – acousto-optic modulator, $T_{Cycle} = 10\ T_R$, CQD – coupled quantum dots, APD – avalanche photodiode. **b**, CQD with a single hole spin in the red (lower transition energy) quantum dot. Two spins in the blue QD inhibit any interaction with the red QD. Lasers are acting on the red QD only. **c**, CQD with one hole spin in each QD and interacting by tunneling-induced exchange interaction, $J$.

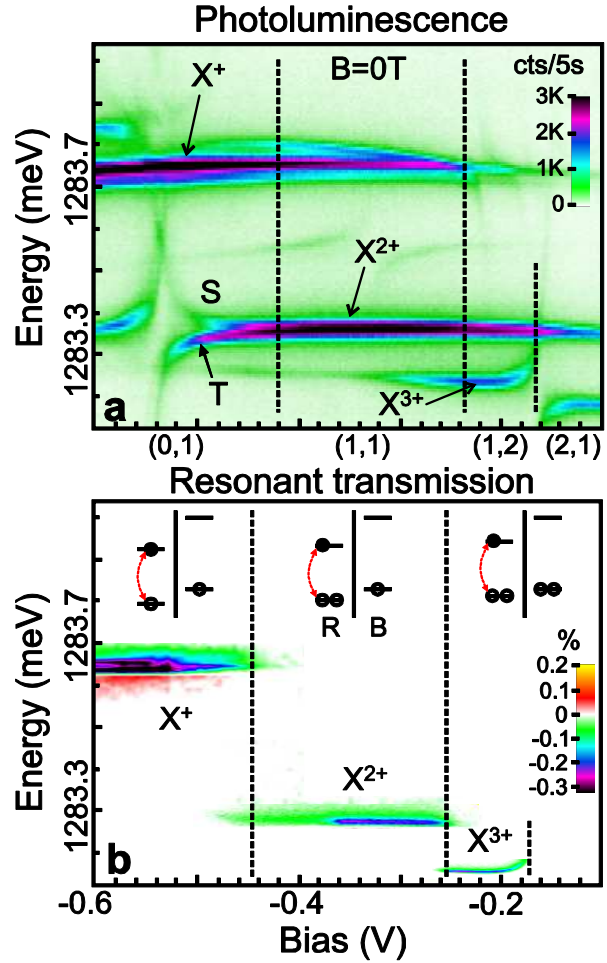

**Figure 2 | Bias-dependent spectroscopy. a**, Photoluminescence spectra of CQD at B = 0T. Emission energy depends on the applied bias showing different exciton charge complexes. The X-shaped pattern on the left gives a unique identification of the $X^{2+} \begin{pmatrix} \updownarrow & 0 \\ \Updownarrow\Downarrow & \Updownarrow \end{pmatrix}$ charge configuration with triplets (*T*) and singlet (*S*) as the ground state. The *S-T* exchange splitting decreases for less negative bias and goes below the resolution of PL measurement (15 µeV). **b**, Bias dependent resonant laser absorption of the red dot. Dashed lines mark the stable regions of different charge configurations. R and B are the red and blue QD, respectively.

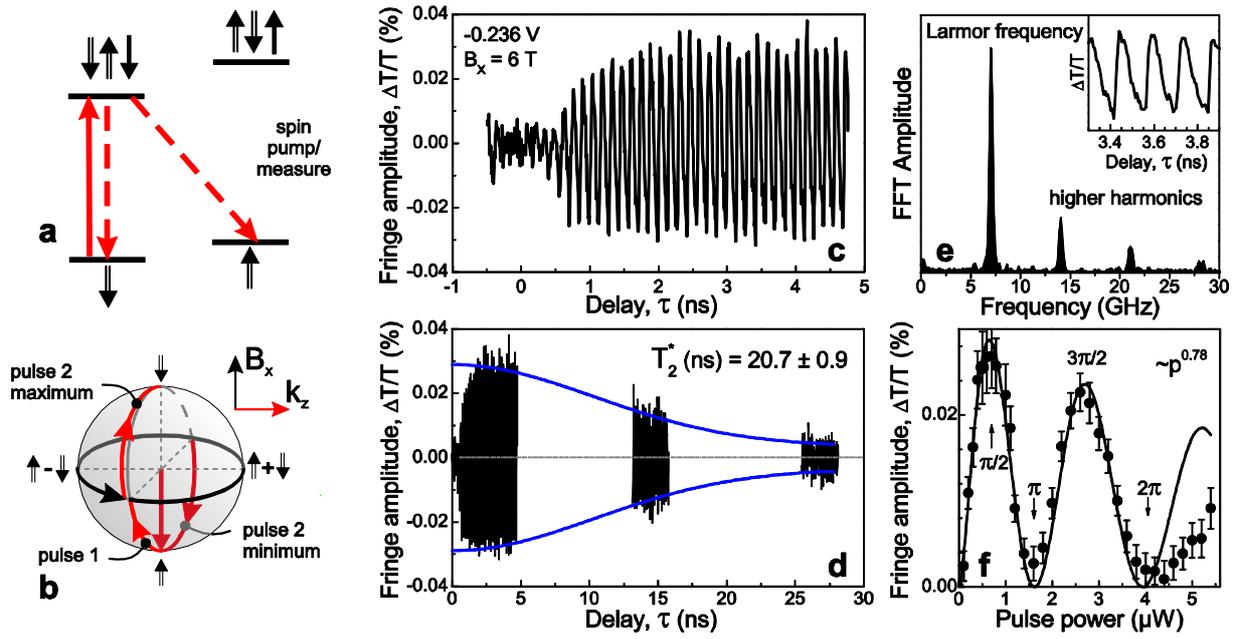

**Figure 3 | Single hole spin control. a**, Λ - level diagram for a single hole spin in transverse magnetic field. The red solid arrow shows the *cw* diode laser. Dashed lines are the recombination paths. **b**, Bloch sphere representation of the hole spin. The magnetic field rotates the spin around the vertical axis labeled $B_x$ (see black arrow around equator), and the optical field rotates the spin around the optical axis labeled $k_z$ (see red arrows moving up the sphere). Depending on the phase evolution, the second pulse brings the spin back to its initial state or to the opposite spin state. **c**, Ramsey fringe experiment with two π/2-pulses. **d**, Same as **c**, but with extended delay. Blue envelope line is the fit with Gaussian decay constant $T_2^*$. **e**, Fourier transform of spectra in **c**. Inset shows the saw-tooth form of the signal that gives rise to the higher harmonics. **f**, Pulse power dependence. Black solid line is the fit using: $\exp(-p/\tau)\sin(\omega p^x)^2$, with x = 0.78. The error bars are obtained from the fits to the Ramsey fringes at each power and represent one standard error.

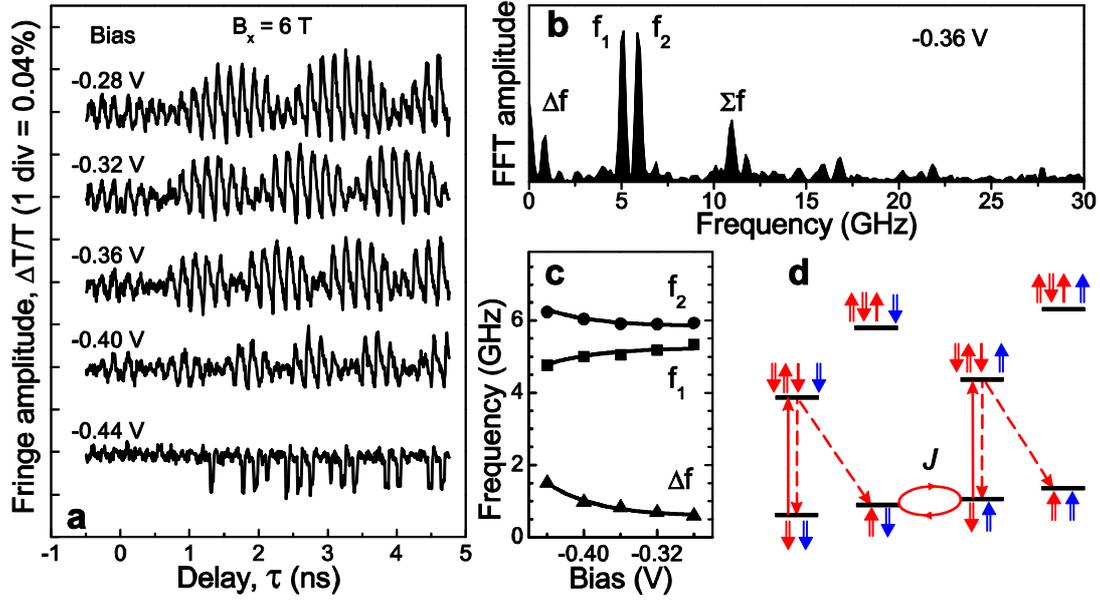

**Figure 4 | Two weakly interacting hole spins. a**, Ramsey fringe experiment for different biases corresponding to different exchange values. **b**, Fourier transform spectrum demonstrating two frequenciesy, $f_1$ and $f_2$, and their sum and difference frequencies caused by the non-linear effects. **c**, Bias dependence of $f_1$, $f_2$, and $\Delta f$ from the Fourier transforms. The lines are calculated using the exchange splitting from the PL in Fig. 2a. **d**, Double $\Lambda$- level diagram for optical transitions of the red QD (red arrows) with exchange (*J*) with the blue QD. Blue dot (shown in blue) is not accessed optically.

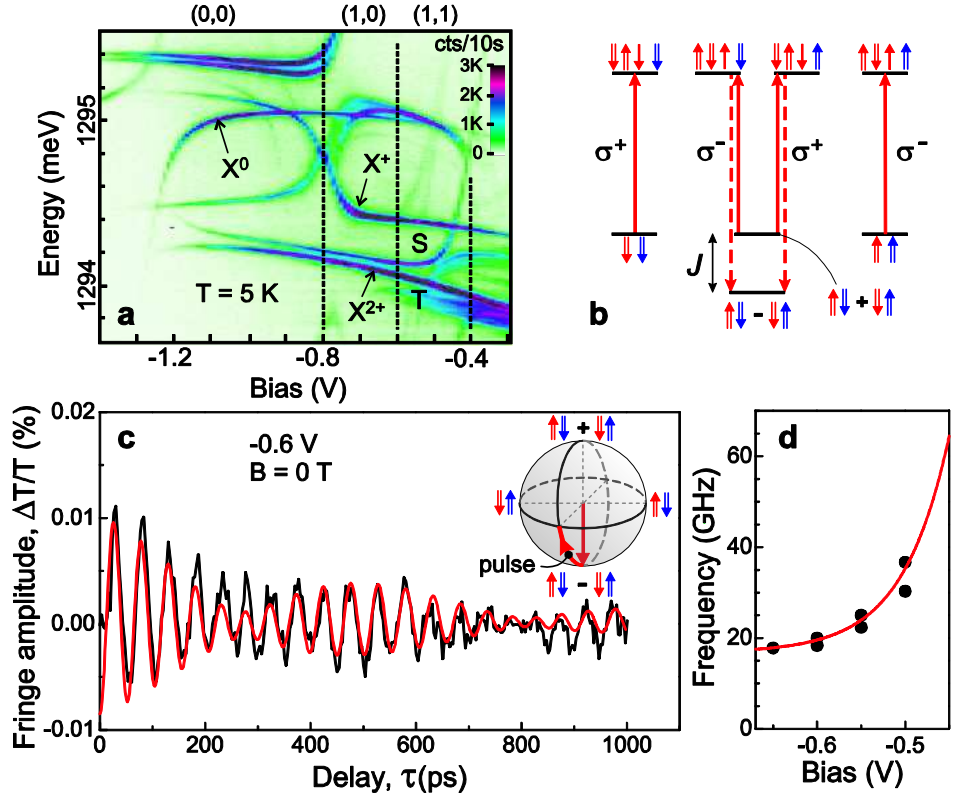

**Figure 5. | Two strongly interacting hole spins. a**, Photoluminescence spectra at B = 0T. The X-shaped pattern corresponds to the $X^+$ charge configuration, with the hole moving between the two QDs. The two lower energy lines correspond to the triplets (T) and singlet (S) of the $X^{2+}$ charge configuration with one hole in each QD. Dashed, vertical lines mark the stable regions of different charge configurations. **b**, Level diagram for optical transitions of the red QD for an isotropic exchange interaction. Double (single) arrows represent hole (electron) spins. **c**, Ramsey fringes at -0.6V. Black curve is experimental data, and the red curve is a model fit. The system is initialized into the singlet by optically pumping out of the triplets. Oscillations are between triplet and singlet states. The inset shows the singlet-triplet Bloch sphere. **d**, Fringe frequencies (solid circles) and singlet-triplet splitting (red line, from exponential fit) vs. applied bias.